\documentstyle[12pt,psfig]{article}
\if@twoside m
\else
\fi
\newlength{\figwidth}
\def\e{\varepsilon}
\def\d{\partial}
\def\o{\omega}
\def\prl{Phys.\ Rev.\ Lett.\ }

\def\ac{adiabatic curvature}

\def\real{{\rm I\kern-.2em R}}
\def\complex{\kern.1em{\raise.47ex\hbox{
	    $\scriptscriptstyle |$}}\kern-.40em{\rm C}}
\def\Box{\hfill\vbox{\hrule height 0.6pt
        \hbox{\vrule width 0.6pt height 1.8ex \kern 1.8ex
                \vrule width 0.6pt}
        \hrule height 0.6pt}}

\newtheorem{op-identity}{Proposition}[section]
\newtheorem{gauge-invariance}{Proposition}[section]
\newtheorem{tls}{Proposition}[section]
\title{Quantum Transport in Molecular Rings and Chains}
\author{J.~E.~Avron and J. Berger\\ Department of Physics, Technion, 32000
Haifa,
Israel}
\begin{document}
\maketitle
\begin{abstract}
We study charge transport driven by deformations in molecular rings 
and chains.  Level crossings and the associated Longuet-Higgins phase 
play a central role in this theory.  In molecular rings a vanishing 
cycle of shears pinching a gap closure leads, generically, to {\em 
diverging} charge transport around the ring.  We call such behavior 
homeopathic.  In an infinite chain such a cycle leads to {\em 
integral} charge transport which is independent of the strength of 
deformation.  In the Jahn-Teller model of a planar molecular ring, 
X$_{p}$, there is a distinguished  cycle in the space of 
uniform shears which keeps the molecule in its manifold of ground 
states and pinches level crossing.  The charge transport in this cycle gives 
information on the derivative of the hopping amplitudes.
\end{abstract}
\newpage
\section{Introduction}\label{I}
Adiabatic transport describes the nondissipating
response of  systems driven by time dependence
\cite{avron,kohmoto,kunz,niu91,seiler91,thouless94}. An attractive feature
of  this theory is the geometrization of linear response theory.  
Its central result is the identification of
transport coefficients with components of the adiabatic curvature. 
Because the theory describes reactive--rather than dissipative--transport, intuition that
comes from dissipative (ohmic) transport is of no real value. For example, it
allows for  charge transport  in materials that are nominally
insulators
\cite{niu94}. The classical counterpart of the theory was studied in
\cite{berry-robbins}.

The focus of our study will be  the role of points of  gap 
closures. This is where one  expects interesting things to happen.  This is also
where one can expect to find universal behavior which is, to a large
extent, system independent. The  basic questions we shall ask is: 
 What distinguishes  a cycle that pinches a gap
closure from a cycle that does not?

The answer to this  question turns out to be related to a question
asked by Longuet-Higgins
\cite{longuet}: Can  one tell if a cycle of Hamiltonians traps  a
point of level crossing? The answer to Longuet-Higgins question is provided by 
looking at what is now known as Berry's phase
\cite{berry84,simon83}. In the special case of
Hamiltonians invariant under time reversal, the phase is $\pm 1$ and is 
known as the Longuet-Higgins phase. A Longuet-Higgins phase of
$-1$ is associated with conic (aka generic) level crossing, and no crossing
has phase
$1$. The Longuet-Higgins phase has observable implications to the
vibrational and rotational spectra of molecules \cite{mead,mt,delcatraz,st}. What we add to
this is that this phase also has observable implication for  electronic
transport. When the cycle of deformation does not trap a gap closure (so 
Longuet-Higgins phase
 is $1$), the charge transport is proportional to the {\em area} of the cycle.  The response
to weak driving is small.  In contrast, a cycle
that traps a generic level crossing, with Longuet-Higgins phase 
$-1$, is characterized by homeopathic response, where the charge transport in a
 molecular ring is {\em inversely} proportional to the length of the cycle. 
In particular,  a
vanishing cause leads to a diverging response. In the case of infinite chains  the behavior
is somewhat different and in some ways more intriguing. It is still true that  there is a weak
response for a tiny cause when the cycle does not trap a gap closure. But, for a
cycle that traps a gap closure, the charge transport is {\em integral},
independent of the strength of the driving.

Charge pumps are quantum systems where
quantized charge transport is driven by a periodic 
modulation \cite{niu90,thouless83}, and have been recently realized 
\cite{pepper}. The  Thouless-Niu pumps  are non-homeopathic: once the
driving becomes  weak the charge transport drops discontinuously to zero. 

Charge transport driven by deformations may be thought of as a generalized version
of piezoelectricity. Piezoelectricity was put in a geometric framework in
\cite{resta,vanderbilt}. In this case, one replaces the
global question --- what is
the charge transport by a full cycle of deformation --- by the local 
question: What is the ratio of current  to the rate of deformation.
Diverging 
piezoelectricity in Hubbard models of perovskites near gap closures was
found in  numerical studies in \cite{ressor}. As we shall explain, near a generic crossing the
ratio of current to rate of deformation is inversely proportional to the distance from the
crossing in an infinite chain and inversely proportional to the square of the distance in
a molecular ring.

The divergence of transport  near gap closures has a geometric interpretation. It
is an analog of an elementary fact  about the curvature of 
surfaces such as spheres and tori. The  average (Gaussian)
curvature $\langle \Omega \rangle$ of a surface with $g$ handles, whose area is
$\e^2$ is, by Gauss-Bonnet,
$$\langle \Omega \rangle= \frac{4\pi\,(1-g)}{\e^2}.$$  
The homeopathic behavior of transport is basically
the  divergence of the (average) curvature as the area, $\e^2$, shrinks to zero. $\e$ is a
measure of the distance from the point of level crossing. In  molecular rings the ratio
of current to rate of deformation is analogous to the Gaussian curvature and diverges like
$\e^{-2}$.  The charge transport in a cycle pinching a gap
closure behaves like a line integral of the curvature and is  of the order of
$1/\e$. 

In the case of infinite  chains with infinitely many 
electrons there is an extra integration over the occupied states
in the Brillouin zone. As a consequence, the
ratio of current to rate of deformation behaves like a line integral of the
curvature and diverges like
$1/\e$.  The
charge transport in a cycle of deformation is then like an area integral of the
curvature. For a $g$ handle surface this gives
$2(1-g)$, independent of its area. The corresponding statement for charge
transport is that a cycle of deformations transports an integer
charge independent of the size of the deformations
(provided it encircles the
point of gap closure).
A model for a charge pump that transports integral charge for infinitesimal cycles
is the insulating infinite  helix described in section \ref{HCP}.

Quantum charge transport driven by deformation has a classical analog in the
theory of elasticity.
$SL(2,\real)$,  the  group of  uniform shears and rotations of
$\real^2$, is generated by
\begin{equation}g_0=\pmatrix{0&1\cr -1&0\cr},\ \
g_1=\pmatrix{1&0\cr 0&-1\cr},\ \
g_2=\pmatrix{0&1\cr 1&0\cr}.
\label{generators}
\end{equation}
 $g_1$ and $g_2$ generate shears, and $g_0$ generates rotations.  
 Since $$ [g_1,g_2]=2g_0, $$ the commutators of infinitesimal shears 
 generate rigid rotation.  Taking an elastic body through a {\em 
 small} cycle of shears results in an overall rigid rotation.  This 
 is, perhaps, the most elementary mechanical analog which mimics the 
 ability of cats to land on their feet by rotating while in free fall 
 \cite{sw89,purcel}.  Using shears to rotate is inefficient: a cycle 
 of deformation of size $\e$ gives a rotation by an angle of order 
 $\e^2$.  The charge transported in such a cycle is also of order 
 $\e^2$ if the cycle does not pinch gap closure.  If it does, the 
 charge is of order $1/\e$.  There appears to be no classical analog 
 of this.

A hallmark of the kind of adiabatic response discussed here is that a 
molecular quantum system can act as an ac to dc converter. For example, 
if one drives a cyclic deformation of the molecule by, say, circularly 
polarized radiation, the response will be a dc current that would give 
rise to a dc magnetic field. It was established 
\cite{Cohen-T} that nuclei, in his case $^3$He, can also respond with a dc 
magnetic field to polarized radiation; in this case this was achieved by 
biasing the population of up and down spins. The piezoelectric response 
of molecular rings leads to the same effect by a very different mechanism.

%%%%%%%%%%%%%%%%%%%%%%%%%%%%%%
\section{An Operator Identity }\label{AT}

In this section we derive the basic identity of adiabatic transport which relates the
transport coefficients with the components of the adiabatic curvature.
This identity is not new
\cite{avron,besb,kohmoto,kunz,niu91,seiler91,thouless94}. The purpose of
this section is to present it as a consequence of an apparently new operator identity.

Let $H=H(x,\phi)$ be a family of self adjoint Hamiltonians that depend 
on two parameters $x$ and $\phi$.  $x$  will stand for a deformation of the 
molecule and  $\phi$ for a magnetic flux that threads a 
molecular ring. 

\begin{figure}[thb]
\centerline{\psfig{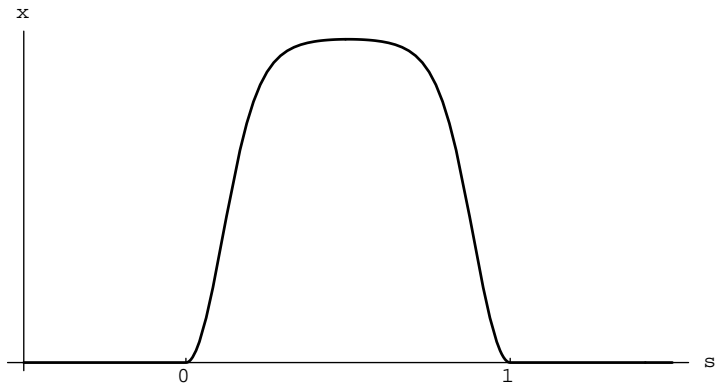}}
\caption{An adiabatic switching of deformation}
\label{switchf}
\end{figure}

 $P=P(x,\phi)$ is a spectral projection for $H$, i.e.  
$PH=HP$, with $P^2=P$.  Suppose that $P$ is finite dimensional and 
smooth.  This is the case, in particular, if $H(x,\phi)$ is a finite 
dimensional matrix, as in the examples we consider.  The parameter $x$ is 
slowly evolving in time in a compact interval, i.e.  $x( t/\tau)$ such 
that $\dot x(s)=0$ for $s<0$ and $s>1$, see fig.  \ref{switchf}.  
$\tau$ is the time scale and the adiabatic limit is $\tau\to\infty$.  
We shall call $s= t/\tau$ scaled-time, and denote by dot derivatives 
with respect to $s$. 
 %%%%%%%%%%%%%%%%%%%%
%%%%%%%%%%%%%%%%%%
 The quantum evolution is described by a unitary $U=U(s,\phi)$  solving the
initial value problem
\begin{equation}i\hbar\,\dot U = \tau\,H\Big(x(s),\phi\Big)U,
\quad U(0,\phi)=1.\label{initial-value}\end{equation}
Let $P_0=P_0(\phi)$ denote a spectral projection for $H\Big(x(0),\phi\Big)$.
$\rho= \rho(s,\phi)= UP_0U^*$ describes the physical evolution of this
projection.   $\rho$ is always a projection, but it is a spectral projection only
if
$H$ is time independent. Since we shall consider Hamiltonians that depend on
time,
$\rho$ is not a spectral projection in general.

The central identity of  adiabatic transport says that
\begin{equation}
\tau\,Tr\Big(\rho\,\partial_\phi H \Big)=
\tau\,\partial_\phi \,Tr (P H)+
\hbar\,Tr\Big(
\Omega_{\phi x}(P)\Big) \, \dot x
+O\left(\frac{1}{\e\tau}\right)
,\label{ad-transport}
\end{equation}
where
\begin{equation}\Omega_{\phi x}(P)=-i\, P\,[\partial_\phi P,\partial_x
P]\,P,\label{ac}
\end{equation} is the $\phi x$ component of the {\em adiabatic
curvature}.
 In the time independent case, $\dot x=0$,
this equation reduces to the Feynman-Hellmann identity. In the time dependent
case Eq.(\ref{ad-transport}) says that the  expectation value of the 
observable $\d_\phi H$ at time
$t$ (in a state that solves the time dependent Schr\"odinger equation and
starts as an eigenstate) is made of two parts: A
{\em persistent response}
$\d_\phi Tr (PH)$ that survives in the $\e\tau\to \infty$ limit, and a
{\em linear response} which is proportional  to the rate of deformation $
\dot x$. The persistent response is leading in the
adiabatic expansion (it is of order zero) while  the linear response is of order
$1/\tau$. In the applications we consider
$\d_\phi H$ is the current operator and $\dot x$ is a rate of deformation. The
\ac\ $\Omega_{\phi x}(P)$  gives, in this case, the
linear dependence of the current on the rate of deformation.

This identity follows from an
{\em operator identity}. For the sake of simplicity we formulate it under 
stronger conditions than necessary:
\begin{op-identity}
Let $P$ be a spectral projection with smooth parametrical dependence on 
$x$ and $\phi$, for a bounded
self-adjoint $H$, so that $P$ is separated by a gap of size larger than
$\e$ from
the rest of the spectrum. Let
$U$ be the solution of the initial value problem, 
Eq.~(\ref{initial-value}), and
$\rho(s,\phi)=U(s,\phi)P_0(\phi)U^*(s,\phi)$ with $P_0$ the spectral projection
at the initial time. Then, the following operator identity holds (for all times
s):
\begin{equation}
\tau\, \rho\,(\partial_\phi H)\,\rho= \tau\, P(\partial_\phi H)P+
 \hbar\,\Omega_{\phi x}(P) \,\dot x
+O\left(\frac{1}{\e\tau}\right).\label{operator-identity}
\end{equation}
\label{op-identity}
\end{op-identity} The point of this identity, Eq~(\ref{operator-identity}), is
that the left hand side requires the solution of an initial value
problem while the right hand side is determined by spectral information alone.

Following \cite{kato58} and \cite{asy}, one introduces an auxiliary evolution,
$U_A=U_A(s,\phi)$,   which respects the spectral splitting with no error.
Respecting spectral splitting means that  $U_A P_0= PU_A$. A generator of such
an evolution is:
\begin{equation} \tau\,H_A=\tau\,H+ i \hbar\,[\dot P,P].\end{equation}
$U_A$ solves the initial value problem:
$$i\hbar\,\dot U_A = \tau\,H_A\Big(x(s),\phi\Big)U_A,
\ \ U_A(0,\phi)=1.$$
The evolution $U_A$ completely suppresses tunneling between the spectral
subspace
$P$ and its orthogonal complement. The adiabatic theorem says that $U_A$ stays
close to $U$:
\begin{equation}
U_A=U+O(1/\e\tau), \ \
U^*\partial_\phi U
=U_A^*\partial_\phi U_A +  O(1/\e\tau).\label{at}\end{equation}
The adiabatic limit is $\tau\e\to\infty$.
 A sufficient condition for the adiabatic theorem to hold  is that $\dot P$ is
smooth and the spectral gap, $\e$, associated with $P$ does not close. For
precise statements and stronger results see \cite{ks,kato58,nenciu}.

From the equation of motion and the chain rule we get:
\begin{equation}\tau\, U^*\, (\partial_\phi H)\, U = i\hbar\, \dot{\hbox\Big(
 U^*\partial_\phi U\Big)}.\label{qdot}\end{equation}
This identity  allows us to
compute transport to order $\tau^{-1}$ with error terms (which we do not
compute) of order
$\tau^{-2}$ while using the adiabatic theorem to lowest order,
Eq.~(\ref{at}). Applying Eq.~(\ref{qdot}) to
the generator of  adiabatic evolution
gives:
\begin{eqnarray}i\hbar\,U_AP_0\ \dot{\Big(U^*_A\d_\phi U_A\Big)}\ P_0 U^*_A&=&\tau
P\,
\Big(\partial_\phi H_A\Big) P=\nonumber \\  \tau P(\partial_\phi
H)P+{i\hbar} P\,[\dot P,\partial_\phi P]\,P
&=&\tau\, P(\partial_\phi H)P+\hbar\,\Omega_{\phi x}(P)\,
\dot x.\end{eqnarray}
From Eq.~(\ref{qdot}), and the adiabatic
theorem:
\begin{eqnarray}
\tau\, \rho\,(\partial_\phi H)\, \rho&=& \tau\,\Big(UP_0\, U^*
\Big)\,(\partial_\phi H)
\,\Big(U\,P_0U^*\Big) =\nonumber\\
{i\hbar}\, UP_0\,\dot{\hbox\Big(U^*\partial_\phi
U\Big) }
\,P_0U^* &=&  i\hbar\,
U_AP_0\,\dot{\hbox\Big(U_A^*\partial_\phi
U_A\Big)}\,P_0U^*_A +O\left(\frac{1}{\e\tau}\right) \nonumber \\
&=&\tau\,P(\partial_\phi H)P +
\hbar\, {\Omega}_{\phi x}(P) \,\dot x+O\left(\frac{1}{\e\tau}\right)
.\nonumber
\end{eqnarray}
We used the adiabatic theorem in the third  step. The other
steps are identities. This completes the proof of the operator 
identity.\hfill$\Box$

Tracing Eq.~(\ref{operator-identity})  gives:
\begin{equation}\tau\, Tr\,\Big(\rho\,\partial_\phi H \Big)= \tau\, Tr(
P\partial_\phi H)+ \hbar\,Tr\Big(
{ \Omega}_{\phi x}(P)\Big) \,\dot x
+O\left(\frac{1}{\e\tau}\right).\nonumber\end{equation}
Since $P\dot{P}P=0$,
\begin{eqnarray}Tr(\partial P)H&=&TrP_\perp (\partial P)PH +TrP(\partial
P)P_\perp H
\nonumber \\
&=&TrPP_\perp (\partial P)H +Tr(\partial
P)P_\perp PH=0,\nonumber\end{eqnarray}
where  $P_\perp=1-P$ is the orthogonal projection on the
complement of $P$. This gives Eq.~(\ref{ad-transport}).

In the adiabatic limit the persistent response dominates except in the case where $\partial_\phi Tr(PH)=0$. In this
case the response in the adiabatic limit is governed by the adiabatic
curvature. This is the case we are interested in.

There are two mechanisms that make the persistent response
vanish:
\begin{enumerate}
\item Time Reversal: Suppose that $H(x,0)$ is time reversal invariant and that 
$\Big(\partial_\phi H\Big)(x,0)$ is odd under time reversal. Then, an
elementary argument shows that $\partial_\phi Tr(PH)(x,0)=0$. There are no
persistent currents in deformed molecules in the absence of external magnetic
field, and provided time reversal is not spontaneously broken.
\item Unitary Families: In the case that the $\phi$ dependence of the
Hamiltonian comes from a unitary transformation, i.e.
\begin{equation}
H(x,\phi)= U(\phi)\, H_0(x)\,U^*(\phi),
\end{equation}
then, clearly $\partial_\phi Tr(PH)(x,0)=0$. \footnote{This  mechanism is responsible for
the absence of persistent currents in infinite chains irrespective of the
question whether time reversal is broken or not.}
\end{enumerate}
Eq.~(\ref{ad-transport})
generalizes to the case where
$x$ is multidimensional. In  the applications we consider  $x$ is two
dimensional and naturally represented as $x\in\complex$.  Assuming no
persistent  response, the adiabatic transport is:
\begin{eqnarray}
 Tr\Big(\rho\,\partial_\phi H \Big)&=&
\frac{ \hbar}{\tau}
\,\, Re \,\Big\{ Tr\Big(
{ \Omega}_{\phi x}(P)\Big)  \dot x\Big\}
+O\left(\frac{1}{\e\tau}\right)\nonumber \\
&=&
\,\hbar\, Re \,\left\{ Tr\Big(
{ \Omega}_{\phi x}(P)\right)\,  \frac {dx}{dt}\Big\}
+O\left(\frac{1}{\e\tau}\right)
.\end{eqnarray}
In a molecular ring the current observable  is
\begin{equation}
I=-{c}\, \frac{\partial H}{\partial \phi},
\end{equation}
where $\phi$ is the flux threading the ring in Gaussian units. It
follows that the charge
$Q$ transported in a cycle in
$x$ space, in the adiabatic limit, with $H(x,0)$ time reversal invariant,
is:
\begin{eqnarray}
Q&=&- {c}\,\oint dt\, Tr\Big(\rho\,\partial_\phi H
\Big)=-{\hbar c}\, Re\oint
\,Tr\Big( { \Omega}_{\phi x}(P)\Big) {dx}\nonumber\\
&=&-{e\Phi_{0}}\, Re\oint
\,Tr\Big( { \Omega}_{\phi x}(P)\Big) {dx}.\label{cad-transport}
\end{eqnarray}
$\Phi_{0}=\frac{\hbar c}{e}$ is the quantum of flux. Note that in 
atomic units where $e=\hbar=1, \ c=137$, the unit of quantum flux is 
a large number. This will be important subsequently, as we shall see.
Eq.~(\ref{cad-transport}) will be our basic tool. Note that the 
factors in front of the adiabatic curvature must be as they are on 
dimensional grounds.

%%%%%%%%%%%%%%%%%%%%%%%%%%%%%%%%%%%%%%%%%%%%%%

\section{Sheared H\"uckel Models}\label{STB}
In the Hamiltonians we consider, $H(x,\phi)$, $x$ denotes a 
deformation and $\phi$ magnetic flux.  The magnetic flux is virtual 
and is only used to define the current operator.  The family $H(x,0)$ 
is time reversal invariant.

By Wigner von-Neumann theorem, level crossings for time reversal 
invariant Hamiltonians are of codimension two.  That is, like a point 
in a plane or a line in three space.  For a cycle of deformation to 
pinch a point of level crossings, $x$ space must be at least two 
dimensional.  The simplest case is when $x$ space is precisely two 
dimensional.

The space of deformation of a molecule with $N$ nuclei is $3(N-2)$ 
dimensional~\footnote{The $3N$ nuclear coordinates include $3$ center 
of mass coordinates and $3$ Euler angles that do not correspond to 
deformations.}.  A two dimensional subspace we shall focus on is the 
space of uniform shears in a plane.

The generators of uniform shears in two dimensions are the $2\times 2$ 
matrices $g_1$ and $g_2$ of Eq.~(\ref{generators}).  Uniform shears 
can be represented by vectors $\vec x$ in $\real^2$, so that the shear 
associated to $\vec x$ is represented by the linear map $1+ 
x_1g_1+x_2g_2$.  It is convenient to use complex notation so $\vec x$ 
is represented by a complex number, $x\in\complex$.  Under the shear 
$x$ a point $d$ in the sheared plane is transformed to $$d\to d+ x\bar 
d.$$
The correspondence between the vector and complex representation is: 
$$\vec x \to x=x_1+ix_2, \ \ \vec d \to d=d_1+id_2, \ \ g_1 \vec d \to 
\bar d, \ \ g_2\vec d \to i\,\bar d.$$
The length of $d$ is stretched according to: \begin{eqnarray} |d|^2 
\to |d +x\bar d|^2&=& {|d|^2 + x\bar d^2 + \bar x d^2+|x|^2|d|^2}.
\label{stretch}
\end{eqnarray}
Suppose, for simplicity, that the hopping amplitude $h_{ij}$ between 
sites $i$ and
$j$  of a planar molecule is a function (typically, an exponential) of the distance
$|d_{ij}|^2$ between the sites:
$h_{ij} = h(|d_{ij}|^2)$. Under a small uniform deformation  the hopping
amplitudes will transform (to linear order) according to
$$h(|d|^2) \to h(|d|^2)+ 2\,{h'(|d|^2)}\, Re\, (x\bar d^2).$$
In particular, the H\"uckel model of such a molecule is, to linear order, of the
form
\begin{equation}
H(x)=H_0+x\,(\partial_{x}H) +\bar x\,(\partial_{\bar x}H) ,  
\end{equation}
where $H_0,\ (\partial_{x}H)$ and $(\partial_{\bar x}H)$ are $N\times N$ matrices.

\begin{figure}[tbh]
\centerline{\psfig{figure=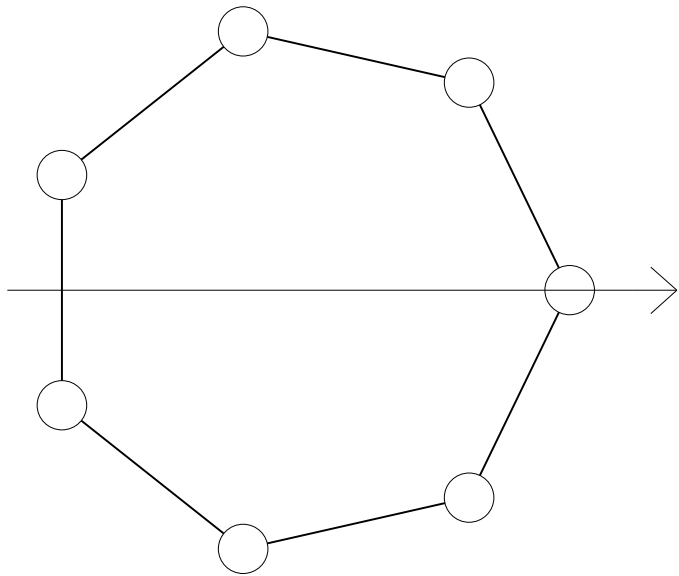,width=\figwidth}}
\caption{Molecular necklace}\label{neckf}
\end{figure}

As an example
 consider the H\"uckel model associated with the sheared
planar molecular  necklace X$_p$ of
$p$ sites.  The tight binding model  of such a
necklace, see Fig.~\ref{neckf},  is
\begin{equation}H(\{h\})=
\pmatrix{0&h(|d_1|^2)&0&.&0&h(|d_{p}|^2)\,\cr
	h(|d_1|^2)&0&h(|d_2|^2)&.&.&0\cr
	 0&h(|d_2|^2)&0&.&.&.\cr
	 .&.&.&.&.&0\cr
	 0&.&.&.&0&h(|d_{p-1}|^2)\cr
h(|d_{p}|^2)&0&.&0&h(|d_{p-1}|^2)&0\cr}.\label{neck}\end{equation}  We
fix the length scale so that the unstrained molecule has $p$ 
equidistant atoms on a circle  of perimeter $p$.  The $d_j$ bond in the unstrained
necklace is a complex number of modulus one:
$$d_j=
\frac{\omega^j-\omega^{j-1}}{|\omega-1|}=i\o^{j-1/2},
$$
with
$\omega=\exp ( i\theta )$,
$\theta= 2\pi/p$ and
$j=1\dots p$.
%%%%%%%%%%%%%%%%%%
%%%%%%%%%%%%%%%%%%
A shear $x$  changes the distances  by
\begin{equation}  
Re \,(\bar d_j^2 x)=- Re \,(\o^{2j-1}\bar x).
\label{elongation}
\end{equation}
This gives the two parameter family of Hamiltonians $H(x,\bar x)$. In particular
\begin{equation}H_0=h(1)\,\pmatrix
	{0&1&0&.&1\cr
	1&0&1&.&0\cr
	 0&1&0&.&0\cr
	 .&.&.&.&.\cr
	 0&.&.&0&1\cr
	  1&.&.&1&0\cr},\ 
\d_{\bar x} H=-h'(1)\,\pmatrix
	{0&\o&0&.&\o^{-1}\cr
	\o&0&\o^3&.&0\cr
	 0&\o^3&0&.&0\cr
	 .&.&.&.&.\cr
	 0&.&.&0&\o^{-3}\cr
	  \o^{-1}&.&0&\o^{-3}&0\cr}.
\label{retake}
\end{equation}

%%%%%%%%%%%%%%%%%%

\section{Adiabatic Curvature Near Crossing}\label{TLS}

The \ac\ near conic crossing was studied in \cite{berry84} in the 
isotropic situation and by \cite{simon83} in the general case.  The 
H\"uckel model of sheared molecular rings turns out to give an 
intermediate situation.  In addition it also leads to certain 
non-conic crossings.  We therefore explore this situation in some 
detail.

Any two level system can be described as
\begin{equation}H_{1/2}(x,\phi)=\bar n \sigma_++ n\,
\sigma_- + n_3\,\sigma_3,\label{spin}\end{equation}
where our convention for Pauli matrices is:
$$
\sigma_+= \pmatrix{0&0\cr
	     1&0\cr
}, \ \ \ \sigma_-= \pmatrix{0&1\cr
	     0&0\cr
} \ \ \ \ \sigma_3=\pmatrix{-1&0\cr0&1\cr}.$$
A formula for the leading behavior of the $x\phi$ component of the 
\ac\ is:
\begin{tls}
Let $H(x,\phi)$ be a Hermitian $2\times 2$ matrix as in Eq.~(\ref{spin}), with
$x\in\complex$ and $\phi\in \real$, such that to leading order in $x$ and
$\phi$ (as $x,\phi\to0$)
\begin{equation}
n\approx gx^m, \  n_3\approx f\phi,\quad f\in\real,\
g\in\complex.\label{parities}\end{equation} Then  to leading order
\begin{equation}
Tr\, { \Omega}_{x\phi}\approx 
\pm\frac{m \,f\vert g\vert^2 \vert x\vert^{2m-2}\bar x}{4i |\vec n|^{3}},\quad 2\vec n
=(n+\bar n, -i(n-\bar n), 2n_3)
\label{leading}
\end{equation}
\label{tls}\end{tls}
 {\em Proof:}
A  result of 
\cite{berry84} is that the curvature is half the spherical angle. In
 complex notation it reads:
\begin{eqnarray}
Tr\, {\bf \Omega}&=&
\pm\frac{1}{2|\vec n|^3}\ \vec n\cdot d\vec n\times d\vec n\nonumber \\
&=&\pm \frac{\bar n\, dn -n\,
d\bar n}{4i\,|\vec n|^3}\wedge dn_3.
\end{eqnarray}
It follows that for a {\em circle} of deformation, $|x|=const$, the
contribution of the
\ac\ to the  charge transport, in the limit $x\to 0$, is:
\begin{equation}
Q_{\pm}\approx\pm\,{e\Phi_{0}\,m\pi \,\,f|g|^2}\,\frac{|x|^{2m}}{|\vec
n|^3}\approx\pm \,e\Phi_{0}{m\pi}\,\,\left\{\begin{array}{ll}\frac{f}{|gx^m|
}&\phi=0;\\
\frac{|g|^2}{f^2}\,\frac{|x|^{2m}}{\phi^3}&\phi\neq 0.\end{array}\right.
\label{circle}\end{equation}
$Q$ diverges as $|x|\to 0$  if, and only if,
$\phi=0$. If $\phi\neq 0$, then $Q$ 
reaches the maximum value $2m\pi e\Phi_0/(3^{3/2}\phi)$ for 
$|gx^m|=\sqrt{2}f\phi$ and vanishes in the limit.

%%%%%%%%%%%%%%%%%%%%%%%%%%%%%%%%

\section{Trimers} \label{HP}    
The  smallest molecular ring is a trimer  X$_3$.
H$_3$ and Na$_3$ are examples. The case of a trimer is special in that the space of
internal coordinates is three dimensional (since a triangle is uniquely determined
by its sides). Three is also the dimension of linear transformations of the
plane: Two shears and one dilation. 

 A canonical choice for the internal coordinates of trimers goes back to 
Jacobi
\cite{mead,mt}. It is instructive to relate these  with
the coordinates associated with uniform shears. We start by reviewing the 
Jacobi coordinates \cite{mead}.

Let $\vec a,\ \vec b, \ \vec c$ be the distances between the 2-3, 3-1 and 1-2
nuclei so that $\vec a+\vec b+\vec c=0$. The Jacobi coordinates
are
\begin{equation}
\vec u = \sqrt{\frac{3}{2}}\,\vec a,\quad \vec v= \sqrt{\frac{1}{2}}\,(\vec b-\vec
c).
\end{equation}    
Equivalently:
\begin{equation}
\vec a = \sqrt{\frac{2}{3}}\,\vec u,\quad \vec b= \sqrt{\frac{1}{2}}\,\vec
v-\sqrt{\frac{1}{6}}\, \vec u,\quad  \vec  c= -\sqrt{\frac{1}{2}}\,\vec
v-\sqrt{\frac{1}{6}}\, \vec u.
\end{equation} 
Evidently, when  $\vec u$ and $\vec v$ are parallel the trimer is a linear 
molecule.
The  scale and shape of the trimer are determined by the three coordinates: 
\begin{eqnarray}
q=\vec u^2+ \vec v^2&=&\vec a ^2 +\vec b ^2+\vec c ^2,\nonumber \\
 qX=2\,\vec u\cdot \vec v,\quad qY&=&\vec u^2-\vec
v^2.
\end{eqnarray} 
Since $|X|,|Y|\le 1$ one can, without loss, set 
\begin{equation}
X= \sin \theta\, \cos\varphi,\quad
Y=\sin\theta\,\sin\varphi,\quad \theta\in[0,\pi/2],\
\varphi\in[0,2\pi).
\label{angl}
\end{equation}
The shape is determined by $X$ and $Y$ alone. This can be seen by
considering the lengths (squared) of the sides:
\begin{eqnarray}
a^2&=&\frac{q}{3}\left(1+X\right),\quad 
  b^2= 
\frac{q}{3} \left(1-
\frac{1}{2}\, X -\sqrt{\frac{3}{4}}\, Y\right),\nonumber \\
 c^2&=& \frac{q}{3} \left(1-
\frac{1}{2}\, X+\sqrt{\frac{3}{4}} \,Y\right).
\label{shape}
\end{eqnarray}

Thinking of $\theta$ and $\varphi$ as the canonical spherical angles, we see that
the equilateral triangle sits at the north pole. The equator corresponds to
degenerate triangles where the trimer is linear.
The locus of isoceles triangles are the three meridians
\begin{equation}
\tan \varphi =0,\quad \tan \varphi =\pm\sqrt{3}.
\end{equation}
And finally, $\varphi \to \pi -\varphi$ reverses orientation. This completes 
the description of the Jacobi coordinates.

The area (squared) of the triangle is proportional to
\begin{equation}
4\,| \vec u\wedge \vec v|^2=q^2\,(1-Y^2-X^2).
\end{equation}
We are interested on triangles which are almost
equilateral, that is, near the north pole.
To linear order in $X$ and $Y$, constant area is the same as constant $q$. 
Comparing Eq.~(\ref{shape}) and Eq.~(\ref{stretch}) we see that, within this
linear regime, an almost equilateral triangle with given coordinates $X$ and
$Y$ is obtained by applying a shear to an equilateral triangle, where the
complex $x$ defined in section \ref{STB} is
%\begin{equation}
$x=\frac{X}{2}+i\frac{Y}{2}$.
%\end{equation}
For this identification we have taken an equilateral
triangle oriented so that $\vec a$ is parallel to the real axis of the
complex plane; rotation of the equilateral triangle by $\varphi_0/2$ in the
complex plane amounts to
multiplication of $x$ by a constant phase $e^{-i\varphi_0}$.
%%%%%%%%%%%%%%%%%%%%%%%%%%%%%%%%%

Consider now the molecular trimer X$_3$ threaded by a fictitious flux, so that the
H\"uckel model
\footnote{The  model   is naturally associated
with three ($\pi$) electrons molecule, e.g. $H_3$ molecule. If the electrons are
non interacting, and spin degeneracy is taken into account, then the ground state
is fully occupied by the two electrons and
the first excited state occupied by one electron. The charge transported (to
leading order in $x$)  is the same as the charge transported by a single
electron in the ground state. The currents of the two other electrons mutually
cancel.} becomes
\begin{equation}H(a,b,c,\phi)=\pmatrix{0&a&\bar \xi c\cr
	   a &0&b\cr
	   \xi c&b&0\cr},
\label{triangle}
\end{equation}
where, with abuse of notation, $a,b,c$ are shorthand for 
$h(\vec a^2),h(\vec b^2),h(\vec c^2)$. The virtual flux is 
\begin{equation}\xi=\exp 2\pi i\left(\frac{\phi}{\Phi_0}\right).
\end{equation}
We chose a
gauge so that the flux is associated with the
$c$-bond.

Since the characteristic equation for $H(a,b,c,\phi)$ is
$$-E^3 +E(a^2+b^2+c^2) + 2abc \cos 2\pi
\left(\frac{\phi}{\Phi_0}\right)=0,$$ eigenvalues crossing occurs at
$$\pm (a^2+b^2+c^2)^{3/2} + 3^{3/2} abc \cos 2\pi
\left(\frac{\phi}{\Phi_0}\right) =0.$$
Since the geometric mean is always less than the
arithmetic mean: $$(a^2+b^2+c^2)^{3/2}\ge  3^{3/2} abc,$$ crossing can only
occur if the flux is an integer or half an integer:
 $\xi =\pm 1$. Moreover since the geometric and
arithmetic mean coincide only if all the elements are identical, the 
locus of crossings is $|a|=|b|=|c|$ and $\xi=\pm 1$.
The simple  eigenvalue is $2a$ (the top state if $a$ is positive) and the
corresponding eigenvector is
$|1\rangle= \frac{1}{\sqrt 3}\,(1,1,1)$. (Recall that $\xi= 1$).  The
degenerate eigenvalue is $-a$ and is two fold degenerate.

\begin{figure}[thb]
\centerline{\psfig{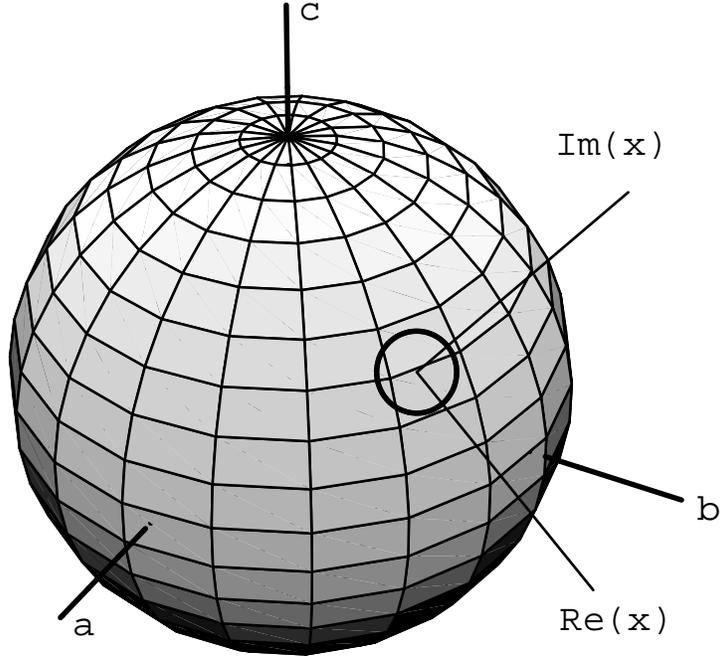}}
\caption{A cycle of deformation}
\label{cycle}
\end{figure}

We consider now a small
cycle of deformations which pinches the line of level crossing, see 
fig. \ref{cycle}.
%%%%%%%%%%%%%%%%%%%% 
%%%%%%%%%%%%%%%%%%%%
Such a cycle is
\begin{eqnarray} a(x)&=&h(1)+h'(1)(\o \bar x + \bar\o x);\nonumber\\  b(x)&=&h(1)+h'(1)
(\bar x +  x);\\ c(x)&=&h(1)+h'(1)(\overline{\o x} + \o x),\nonumber
\label{maslul}
\end{eqnarray}
where $\o$ is a cube root of unity and 
$x\in\complex$ runs on a small circle in the complex
plane surrounding the origin. The phase of $x$ has been chosen as in section 
\ref{STB}.

We are interested in the amount of charge moving around the triangle when 
the molecule undergoes this cycle of deformations that encircles the line of
gap closures. In the case of a single electron, this charge can be interpreted
as the total number of turns around the molecule that the electron makes. 

The expansion of the Hamiltonian near zero flux and
 near the equilateral triangle takes the form :
\begin{equation}
H(x,\phi)=H_0 +h'(1)\,\big(x\,V_x+\overline{x\,V_x}\big)+ 2\pi\, h(1) 
\left(\frac{\phi}{\Phi_0}\right)\, 
V_\phi , 
\end{equation}
where
\begin{eqnarray}
V_x=
 \pmatrix{0&\bar \o&\o\cr
	   \bar\o &0&1\cr
	   \o&1&0\cr} &;& \ \ \
V_\phi=i\, \pmatrix{0&0&-1\, \cr
	   0 &0&0\cr
	   1 &0&0\cr}.\nonumber
\end{eqnarray}
The degeneracy
splits in first order of perturbation theory, both in $\phi$ and in $x$.  The
local behavior near crossing is given by the
$2\times 2$  matrix as in Eq.~(\ref{spin}) with
\begin{equation}
f=\frac{2\pi h(1)}{\Phi_0\sqrt 3},\quad g=2\bar\omega h'(1),\quad
m=1.\end{equation} 
 For  a circular orbit, $|x|=const$., the total charges transported per
cycle in  the  almost crossing state is given by Eq.~(\ref{circle}). If
there is no external magnetic field, so the flux $\phi=0$, one finds:
\begin{equation}
 Q= \pm \,e \left(\frac{\pi^{2}}{\sqrt{3}}\right)\,\left(\frac{
h(1)}{\,h'(1)}\right)\,\frac{1}{|x|}.
\end{equation}
The divergence of the charge transport near conic crossing is 
universal for conic crossing.  It describes the remarkable fact that 
the smaller the cycle that pinches the degeneracy, the more charge it 
transports. The overall constant is inversely proportional to the 
logarithmic derivative of the hopping amplitude. 

The current is {\em not} large.  Only the ratio of current to the rate 
of driving is large.  As the circle is shrunk, the rate of driving 
must also decrease in order for the adiabatic theory to apply.

Driving a system by an infinitesimal cycle that pinches a 
crossing may or may not be an easy thing to do.  In molecular 
rings, X$_{p}$, this may well be a difficult thing to accomplish, since  
the Jahn 
Teller instability fixes a pinching cycle, or finite radius, which is 
the ground state manifold of the molecule. This point is discussed in 
more detail in the last section.

%%%%%%%%%%%%%%%%%%%%%%%%%%%%%%%%%%%
\section{Sheared Molecular Necklaces}\label{SN}
 The study of charge transport in necklaces of $p$ atoms with zero 
 threading flux requires analysis of the gap closures for $\xi=1$.  We 
 take the Hamiltonian (\ref{retake}), with flux dependence given by 
 the obvious generalization of (\ref{triangle}).  %%%%%%%%%%%%%%
\begin{figure}[tbh]
\centerline{\psfig{figure=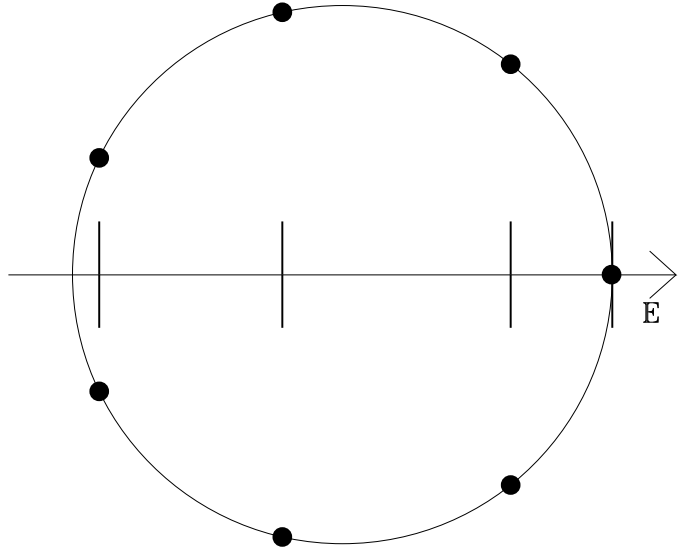,width=\figwidth}}
\caption{Eigenvalues and the roots of unity}
\label{roots}\end{figure}
%%%%%%%%%%%%%%
The eigenvalues of the unstrained
molecule are:
$$E_m=2\cos (m\theta), \ \ m=0,\dots ,\frac{p-1}{2}, \ \ 
\theta=\frac{2\pi}{p}.$$
For simplicity, we deal here only with the case that $p$ is odd.
Note that the energies are in decreasing order, fig \ref{roots}. All energies,
except
$m=0$ are doubly degenerate so that the projection on the m-th subspace is
spanned by
$|\pm m\rangle$ where
$$|m\rangle=\frac{1}{\sqrt p}
\pmatrix{\o^m\cr \o^{2m}\cr\o^{3m}\cr\dots\cr 1},\ \
\ m=0,\pm 1,\dots,\pm\frac{p-1}{2}.$$
One finds
$$\sqrt p \,(\d_{\bar x} H)
|m\rangle=-h'(1)(\bar\o^{m+1}+\o^{m+1})\ \pmatrix{\o^m\cr
				\o^{2m+2}\cr
				 \o^{3m+4}\cr
....\cr
\o^{-2} };
\ \ \sqrt p\,(\d_{\phi} H) |m\rangle=\frac{2\pi ih(1)}{\Phi_0}\pmatrix{-1\cr
				0\cr
....\cr
0\cr
\o^{m}}.$$
 $\d_{\bar x}H$ is  a double right shift:
\begin{eqnarray}\langle m |\d_{\bar x}H|
{m+k}\rangle&=&-h'(1)\,\frac{
\bar\o^{m+k+1}+\o^{m+k+1}}{p}\ (\o^k+\o^{2k+2} +\o^{3k+4}+\dots )\nonumber \\
&=&
-h'(1)\,\delta (k+2)\,\bar\o^2 \,(\bar\o^{m-1}+\o^{m-1}).
\end{eqnarray}
 Similarly,
$$\langle m |\d_{\phi}H|k\rangle=
2\pi ih(1)\,
\frac{\o^k-\bar\o^m}{p\Phi_0}=-\frac{4\pi h(1)}{p\Phi_0}\,\o^{\frac{k-m}{2}}
\sin (\frac{k+m}{2}\theta)
$$

\begin{figure}[tbh]
\centerline{\psfig{figure=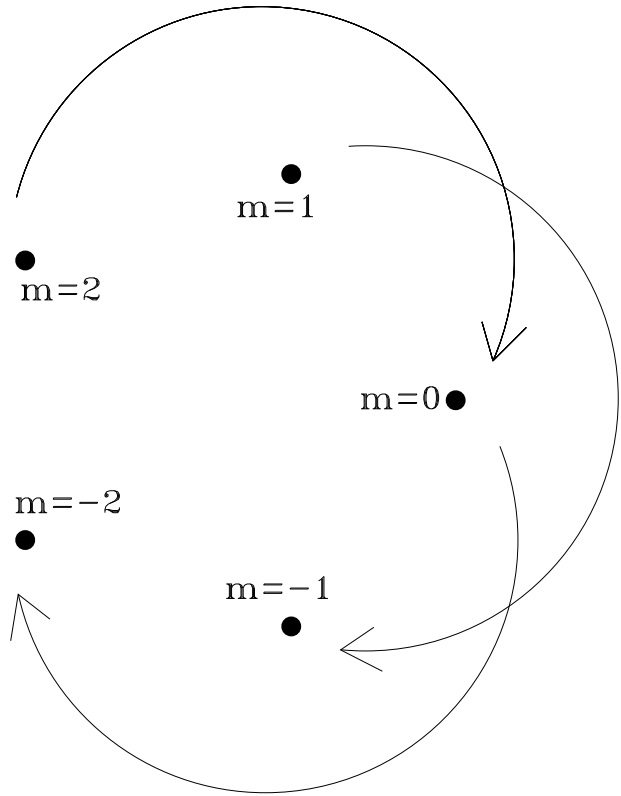,width=\figwidth}}
\caption{Coupling of states by the perturbation}
\label{hop}
\end{figure}

The two dimensional subspace associated with $|\pm m\rangle $, for   
$m=1,\dots,\frac{p-1}{2}$, is split in first order of perturbation theory in the
$\phi$ variable. In contrast, the split is in order
$m$ of perturbation theory in the
$x$ variable. All we need to compute
is the $2\times 2$ matrix describing the crossing, and
in the case $\xi=1$, see fig \ref{hop}, we find (up to an overall scale):
\begin{eqnarray} H_{1/2}=&2(-h'(1))^m&
\left( \prod_{k=1}^{m-1} \frac{ \cos (-m+ 2k-1)\theta }{\cos m \theta
-\cos
 (m-2k)\theta}\right) \, \cos(m-1)\theta\ \left( \overline{(\o^2 x)}^m
\sigma_+ + (\o^2 x)^m\sigma_-\right)\nonumber \\
&-&\frac{4\pi h(1)}{p\Phi_0}\,\sin(m \theta)\,\sigma_3\,\phi.\label{pauli}
\end{eqnarray}
%%%%%%%%%%%%%%
%%%%%%%%%%%%%
This puts us in the framework of proposition \ref{tls} with explicit 
values for $g$, $m$ and $ f$.  Note that as far as degenerate 
perturbation theory is concerned, there are also other terms that 
split the degeneracy, like $x\phi$.  However all these are irrelevant 
for computing curvature since $g$, $m$ and $f$ are all we need in 
Eq.~(\ref{leading}).  The transport is given by Eq~(\ref{circle}).

We want to conclude this section with two cautionary remarks about 
the scope of validity of the results of this section.  The moral of both of 
these is that the results for conic crossings have a wider scope of 
validity than the results for non-conic crossings.
\begin{enumerate}

\item The theory of deformed tight binding Hamiltonians was developed 
to first order in $x$.  For gaps that open to first order, that is for 
conic crossings, such a perturbation expansion is consistent.  
However, for gaps that open in higher orders, one should have 
considered  sheared tight binding models 
valid to higher orders in $x$.
To lowest order, Eq.~(\ref{pauli}) may or may not remain applicable,
depending on whether there exist paths that connect $|m\rangle$ to
$|-m\rangle$ more efficiently than in Fig.~\ref{hop}. 
 \item We have restricted the space of deformations to uniform shears.  When 
other deformations are also possible, as is the case for planar 
molecules X$_{p}$ with $p\ge 4$, these cannot be 
neglected: it is always possible to build a deformation that will open 
the gap in first order and will therefore provide the lowest order 
perturbation.
\end{enumerate}
%%%%%%%%%%%%%%%%%%%%%%%%%%%%%%%%%%%%%%%%%%%%%
%%%%%%%%%%%
\section{Crossing Bands and Direct Integrals}\label{CB}
In this section we describe a version of the equation for adiabatic
transport, Eq~(\ref{ad-transport}),  for systems described by direct integrals.
This is the case for infinite periodic chains of non-interacting electrons that
admit Bloch decomposition \cite{grossmann}. From a physical point of view there
are two important consequences to this extra structure. The first is that
charge transport becomes integral. The second, and related, is that the
formulas for charge transport, which for finite molecular rings involved
non-universal factors (the logarithmic derivative of the hopping
amplitudes), become truly universal. Studies of Chern numbers in the context
of charge transport have been made in
\cite{bbg,kunz,thouless83,kreft-seiler,novikov,montambaux,tan,wilkinson,yano}.

 Consider the Hamiltonian  represented by a direct integral (over the torus):
$$H_B(x) = \int_\oplus H(x,\phi )\, \frac{d\phi}{2\pi}.$$
$H(x,\phi )$ acts on a fixed Hilbert space, say ${\cal H}$.  In this and 
the following section $\phi$ will stand for Bloch momentum. A spectral 
projection on an energy band is
$$P_B(x) = \int_\oplus P(x,\phi )\, \frac{d\phi}{2\pi}.$$  $P(x,\phi)$, the
spectral projection for $H(x,\phi )$,  is smooth and periodic in
$\phi$. Consider the  observable
$$\Big(\partial_\phi H_B\Big)(x)=
\int_\oplus
\partial_\phi H(x,\phi )\,
\frac{d\phi }{2\pi}.$$
This is interpreted as current: It
is made up from velocity $\partial_\phi  H(x,\phi )$ and density
$\frac{d\phi }{2\pi}$. In this case, $\phi $ runs on the unit circle. The
persistent response for infinitely many non interacting fermions filling the
band  vanishes by periodicity:
\begin{equation}
Tr\,\Big(P_B\d_\phi H_B\Big)=\int_{-\pi}^\pi \frac{d\phi }{2\pi}
\,\partial_\phi
\, Tr\, (P H)=0.
\end{equation}

{\em Nearly Crossing Bands}:  The question we address is what can one say
about ``axial'' integrals of the \ac\
\begin{equation}\int_{-\pi}^\pi \frac{d\phi }{2\pi}Tr\, \Omega_{\phi
x}\Big(P(x,\phi)\Big) =-{i}\,\int_{-\pi}^\pi
\frac{d\phi }{2\pi}\, Tr\,\Big( P(x,\phi )
\,[\d_\phi P(x,\phi ),\d_x P(x,\phi )]\Big),\label{bands}\end{equation}
in the case of nearly crossing bands. More precisely, suppose that $x=0$ is a
band closure for some value of $\phi$.
 What is the analog of Eq.~(\ref{circle}) in
this case?

 Suppose that
$P(x,\phi )$ is a projection in
$\complex^2$ associated with the two level Hamiltonian Eq.~(\ref{spin}),  with
$n(x,\phi)$ and $n_3(x,\bar x,\phi)$ as in proposition \ref{tls} and we 
set the origin of
$\phi$ space at the zero of
$n$ and $n_3$, assuming approach to zero as in Eq.~(\ref{parities}). In 
general, 
there may be several such zero points. The contributions of these behave
additively and
it is therefore enough to consider one.   Since the
$d\phi $ integral is dominated by the singularity:
\begin{eqnarray}\int_{-\pi}^\pi\frac{d\phi }{2\pi}\, Tr\, \Omega_{\phi
x}\Big(P(x,\phi )\Big)&=&\pm
\int_{-\pi}^\pi \
\frac{f}{4i\,(|n|^2 +n_3^2)^{3/2}}  \bar n
\frac{\d n}{\d x} \frac{
d\phi}{2\pi}
\nonumber \\
&=&\pm \frac{m|g|^2}{8\pi i} |x|^{2m-2}
\bar x \,\int_{-\infty}^\infty \
\frac{f\,d\phi }{(|gx^m|^2 +|f\,\phi|^2)^{3/2}}
 +O(1)
\nonumber \\
&=&\pm \frac{m}{8\pi i|x|^2}
\bar x \,\int_{-\infty}^\infty \
\frac{d\phi }{(1 +\phi ^2)^{3/2}}
 +O(1)
\nonumber \\
&=&\pm \frac{m}{4\pi ix}
 +O(1).
\end{eqnarray}
A similar formula holds for $\Omega_{\phi\bar x}(P)$. Now the divergence is like
$x^{-1}$ and the power is independent of the order of crossing. The order shows
up as a linear factor.

\begin{figure}[tbh]
\centerline{\psfig{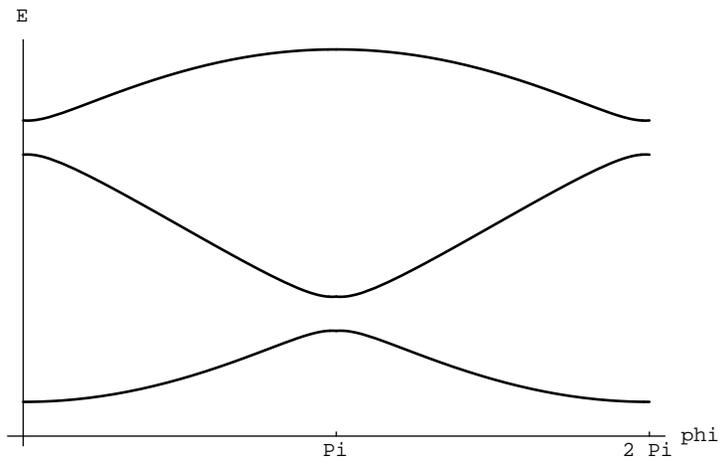}}
\caption{Top and bottom gap closures}\label{tb}
\end{figure}

In the  case that there are several such points the adiabatic response collects
the contribution from all  of them. These come with one
sign for the top gap closure and the opposite sign for the bottom gap closure,
fig. \ref{tb}.
%%%%%

For fixed Fermi energy in a gap, with one point of gap closure, the total
charge transport by a loop\footnote{Not necessarily a circle.} encircling a
gap closure,  is precisely :
\begin{equation}Q= \mp m .\label{gap}\end{equation}

Homeopathy  means that  integral charge is transported
even if the deformation is minuscule.

%%%%%%%%%%%%%%%%%%%%%%%%
\section{Homeopathic Charge Pump} \label{HCP}
%%%%%%%%%%%%%%%%%%%%
\begin{figure}[tbh]
\centerline{\psfig{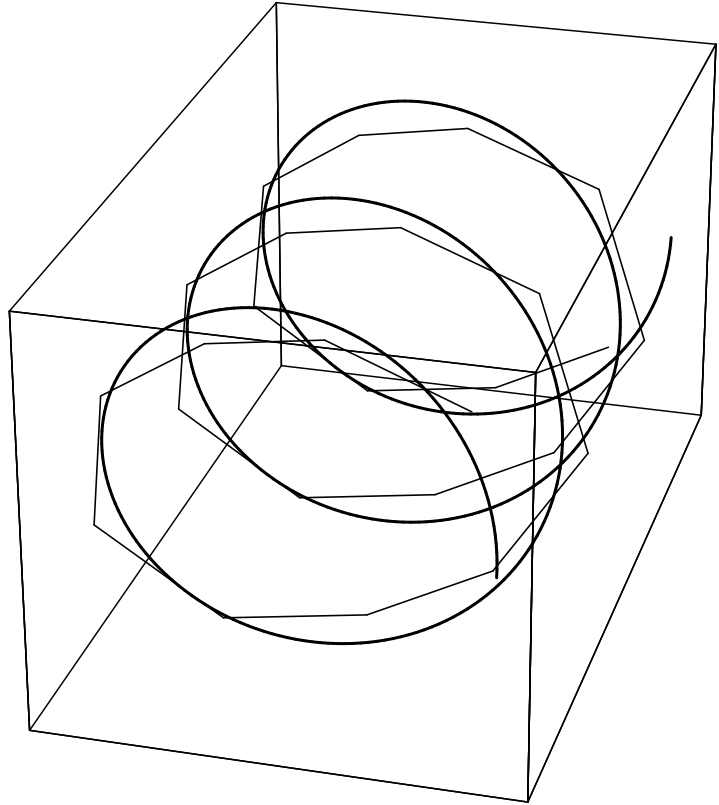}}
\caption{Shearing a helix in the transverse plane}
\label{td}\end{figure}
%%%%%%%%%%%%%%%%%%%
Consider an infinitely long helical chain: A single pitch of the helix 
is made of an open circle with $p$ atoms per pitch.  The cross section 
of the helix is a circle and the unstrained atoms on the circle are 
equidistributed.  Of course, there is no reason in the world why a 
chain of identical atoms would want to form a helix.  So we may 
imagine that this structure is enforced by an external constraint. 
 We shall consider uniform shears of the helix 
in a plane perpendicular to the axis of the helix.

A tight binding model of such a helix is described by a $p\times p$ 
matrix $H(x,\phi)$, where $x\in\complex$ is the deformation and $x=0$ 
is the undeformed helix.  $\phi$ stands for the Bloch momentum, 
traditionally denoted by $k$.  $H(x,\phi)$ is identical to the tight 
binding (H\"uckel) Hamiltonian of a necklace of $p$ atoms enclosing a 
flux $\phi$.  Such Hamiltonians were analyzed in section \ref{SN}.

Consider a quantum system of
infinitely many (non interacting, spinless) electrons on such a helix so that
there are $q$ electrons per $p$ atoms.
The unstrained chain, $x=0$,  has
continuous spectrum  in the interval
$[-2,2]$ without any gaps. It is convenient to think of this as a collection of
$p$ bands with closed gaps:
$$ [-2, 2 \cos (\pi (1-1/p))],\ [2 \cos (\pi (1-1/p)), 2 \cos (\pi 
(1-2/p))],\ \dots  [2\cos(\pi /p), 2].$$
Shearing the chain in the plane perpendicular to the chain
axis, see fig. \ref{td}, will break the spectrum into $p$ bands, so that the
lowest $q$ bands will be occupied.
Consider now a cycle of deformations $|x|=\e$ which avoids the origin. Such a
cycle pinches the gap closure at $x=0$. (See  section \ref{STB}.) Collecting
results from the sections below gives the following result:

{\em A small cycle of shears in the plane perpendicular to the axis of 
an infinite helix, with pitch $p$ ($p$ odd), transports integer (non 
zero) charges from $-\infty$ to $\infty$ even if the orbit is 
arbitrarily small (but not zero).  The amount of transported charge is 
$q/2$ if $q$, the number of electrons per pitch of the helix, is 
even.  If $q$ is odd, then the charge transport is $-(p-q)/2$.  }

These results can be seen as follows. By Floquet theory, the
band edges are determined by $H(x,0)$ and $H(x,\pi )$.  The two
cases are simply related when $p$ is odd since  the Hamiltonian, 
Eq.~(\ref{neck}), satisfies
\begin{equation}H(\{h\},0)U= -UH(\{h\},\pi),\label{as}\end{equation}
with a fixed unitary
$$U_{ij}=(-)^j\, \delta_{ij}.$$
Using this, the spectral  and  geometric results for the zero flux of section
\ref{SN} can be easily translated to the $\phi=\pi$ case.

 The lowest gap near $ 2 \cos (\theta (p-1)/2)$  opens in $(p-1)/2$-th 
order of
degenerate perturbation theory. By symmetry, Eq.~(\ref{as}), this then also
holds
for the top gap. The second gap from the bottom opens, likewise, as the
second
gap from the top.  This opens in first order of degenerate perturbation 
theory,
etc. This is explained in fig. \ref{hop}. The general rule  which gives
the order
in which a gap opens depends on the parity of the gap. The results are
summarized in  a table that also gives the Chern number of the gap:
\begin{center}
\begin{tabular}{||l|l|r||} \hline
Gap &Order & Chern \\ \hline
2j&j& j\\ \hline
2j+1&$\frac{p-1}{2}-j$& $-\frac{p-1}{2}+j$  \\  \hline
\end{tabular}
\end{center}

%%%%%%%%%%%%%%%%%%%%%%%%%%%%%%%%%%%
%%%%%%%%%%%%%%%%%%%%%%%%%%%%%%%%%%%%%%%%%%%%%

\section{Elastic and Magnetic Jahn Teller Instabilities }
\label{JT} 
The analysis carried out so far assumed that  the deformation $x$ and the 
flux
$\phi$ are externally controlled parameters which can be varied at will. In
reality, both are dynamic variables with their own equations of motion. The
question one needs to consider is how this may affect transport.

For example, in our analysis it is important that time reversal is not 
spontaneously broken and the flux $\phi$ was a virtual flux.  The 
dynamical equations could, in principle,  force $\phi\neq 0$. If this 
was to happen then homeopathic response is censored.

Even for a molecular trimer X$_3$ (in three dimensions), the full 
dynamics is a formidable problem which brings out all the intricacies 
of the three body problem \cite{mead,mt,englman,bersuk}.  Similarly, the full 
dynamics of $\phi$ is determined by QED.

The static problem is much simpler. It is determined 
by an energy functional whose minimizers are classical equilibrium 
configurations. For the nuclear coordinates the minimizer gives the ground 
state configurations and for the flux it gives Amp\`ere's law.
Such a  functional is :
\begin{equation}
E(x,\phi) +Q(x)+\frac{1}{2L}\phi^2.
\label{jtme}
\end{equation} 
Here $E(x,\phi)=E(x,\phi+\Phi_{0})$ is an eigenvalue of say the 
H\"uckel Hamiltonian, $Q(x)$ is a classical potential energy and  
$L$ is a geometric constant ($L/c^2$ is the self-inductance
of the molecule).  Note that 
the period of $E(x,\phi)$ in $\phi$ is large since in atomic units 
$\Phi_0=c=137$ is large. 

The minimizer of this functional with respect to the flux  gives
the analog of Ampere's equation (in Gaussian units) 
\begin{equation}
0=\frac{\partial E(x,\phi)}{\partial \phi}+
\frac{\phi}{L}= -\frac{1}{c} \, I +\frac{\phi}{L}.
\end{equation} 

Let us now specialize to the case of planar molecular rings X$_p$ and let 
us assume that the elastic energy $Q(x)$ depends only on the distances $|d_j|$ 
between neighboring atoms.
From Eq.~(\ref{elongation}) the elastic energy of a molecular necklace 
X$_p$, in the Harmonic approximation, is  an isotropic function of $x$:
\begin{eqnarray}
Q(x)&=& \frac{K}{2} \sum (|d_j(x)|-|d_j(0)|)^2=\frac{K}{8} \sum 
(\o^{2j-1}\bar x+{\bar\o}^{2j-1}x)^2
\nonumber\\ &=& \frac{pK}{4} |x|^2,
\end{eqnarray}
where $K$ is the ``spring constant".
From Eq.~(\ref{pauli}), the electronic energy surface of the one-electron 
state near crossing is also isotropic in $x$, and
the energy functional is of the form:
\begin{equation}
\pm \sqrt{|g|^2|x|^{2m}+ (f \phi)^2} +\frac{pK}{4} |x|^2+\frac{1}{2L}\phi^2.
\end{equation} 
In this functional $g, K, L$ are all of order one but $f$ is a small 
number of order $1/\Phi_{0}$.

It is convenient to use rescaled variables
\begin{equation}
y=x^{m}, \quad  \varphi= \phi/\Phi_{0}.
\end{equation} 
In these variables the energy functional is
\begin{equation}
\pm \sqrt{|g|^2|y|^{2}+ (f\Phi_{0}) \varphi^2} +\frac{pK}{4} 
|y|^{2/m}+\frac{\Phi_{0}^2}{2L}\varphi^2.
\label{rescale}
\end{equation} 
Here $g,K,L$ and $f\Phi_{0}$ are all of order one in atomic units,
while $\Phi_{0}=137$.

The minimizers of this functional are as follows:
\begin{enumerate}
\item With the positive sign for the square root the minimizer is the 
trivial solution  $\phi=\varphi=y=x=0$.
\item With the negative sign for the square root and conic crossing, $m=1$,  the 
minimizer has $\phi=\varphi=0$ and 
$|x|=\frac{2|g|}{pK}=\frac{4|h'(1)|}{pK}$. $\varphi=0$ 
because its ``spring constant'' is stiff, of order $10^{4}$.
This solution is the classical Jahn Teller instability. Because the 
energy functional is isotropic, the minimizer is a circle. The charge 
transported in this cycle is
\begin{equation}
Q=\pm\,\frac{\pi^2 e Kh(1)\sin(2\pi/p)}{2\left[h'(1)\right]^2}\,.
\end{equation} 
\item With the negative sign for the square root and 
non-conic crossing, $m\ge 2$, the stiff term in (\ref{rescale}) is now the 
elastic 
spring, which is not harmonic.  Therefore the minimizer is $x=0$ and 
$|\phi|=L|f|$.  This is a magnetic Jahn Teller instability, where there 
is spontaneous breaking of time reversal.  Time reversal is only 
weakly broken because $f$ is small: the spontaneous flux is of the 
order of $10^{-4}\Phi_0$.  As we have already 
stressed in a previous section, conclusions drawn from sheared 
H\"uckel models in situations  where the crossing is not conic, are 
fragile and depend sensitively on perturbations. This may be a reason 
why magnetic Jahn Teller has not been observed.
\end{enumerate}

%%%%%%%%%%%%%
\begin{figure}[tbh]
\centerline{\psfig{figure=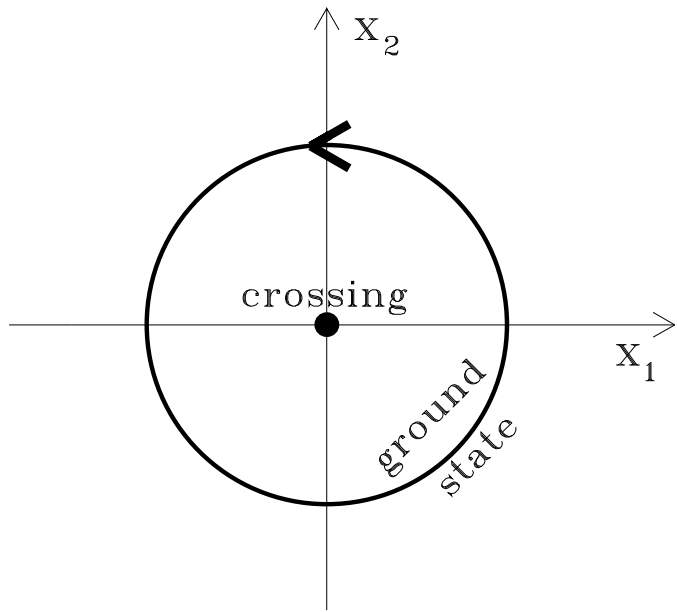,width=\figwidth}}
\caption{The preferred cycle of deformation.} \label{equi}\end{figure}
%%%%%%%%%%%%%

%%%%%%%%%%%%%%%%%%%%%%%%%

%%%%%%%%%%%%%%%%%%%%%%%%%%%%%%%%%%%%%%%%%%%

\section*{Acknowledgments}

We thank  R. Englman, Y.~Last, L.~Sadun, B.~Shapira,
U.~Sivan and P.~Zograf for discussions. This work was partially supported by a
grant from the Israel Academy of Sciences, the Deutsche Forschungsgemeinschaft
under SFB~288, and by the Fund for Promotion of Research at the Technion.
%%%%%%%%%%%%%%%%%%%%%%%%%%%%%%%%%%%%%%%%%%%


\begin{thebibliography}{article}
\bibitem[Avron 1995]{avron}  J.E.~Avron, {\it Adiabatic Quantum Transport}, Les
Houches, E.~Akkermans, G.~Montambaux, J.-L.~Pichard and J. Zinn-Justin eds., 
Elsevier Science.

\bibitem[Avron, Berger and Last 1997]{abl} J.E.~Avron, J.~Berger and Y.~Last, {\it
Piezoelectricity: Quantized  Transport Driven by Adiabatic Deformations}, Phys.\
Rev.\ Lett.\ {\bf 78}, 511-514.

\bibitem[Avron and Sadun 1991]{as} J.E.~Avron and L.~Sadun, {\it
Adiabatic quantum transport in networks with macroscopic component},
Ann.~Phys.~{\bf 206}, 440-493. 

\bibitem[Avron, Seiler and Yaffe 1987]{asy} J. E. Avron, R. Seiler and L. G.
Yaffe, {\it Adiabatic theorems
 and applications to the quantum Hall effect}, { Comm.\ Math.\ Phys. \ }{\bf
110}, 33--49 (Erratum: { Comm.\ Math.\ Phys. \ }{\bf 153} (1993)
649-650).

\bibitem[Avron, Seiler and Zograf 1995]{asz}  J.E.~Avron, R.~Seiler and P.~Zograf,
{\it  Viscosity of Quantum Hall Fluids}, Phys.\ Rev.\ Lett.\ {\bf 75} 697-700.

\bibitem[Bellissard, Bovier and Ghez 1991]{bbg} J.~Bellissard, 
A.~Bovier and J.M.~Ghez, {\it Gap labeling theorems for discrete one dimensional
Schr\"odinger operators},  Rev.\ Math.\ Phys.\ {\bf 4}, 1-37.


\bibitem[Bellissard, van Elst and Schulz-Baldes 1994]{besb} J.~Bellissard,
A.~van Elst and H.~Schulz-Baldes, {\it The non commutative geometry of the
Quantum Hall effect}, J.\ Math.\ Phys.\ {\bf 35}, 5373-5451.

\bibitem[Berry 1984]{berry84} M.V.~Berry {\it Quantal phase factors accompanying
adiabatic changes}, Proc.\ Roy.\ Soc.\ London A {\bf 392} 45-57.
\bibitem[Berry 1989]{berry89} M.V.~Berry {\it The
quantum phase: Five years after}, in
 {\it Geometric Phases in Physics}, A.~Shapere and F.~Wilczek, Eds., World
Scientific, Singapore.

\bibitem[Berry and Robbins 1993]{berry-robbins} M.V.~Berry and J.M.~Robbins, {\it
Classical Geometric Forces of Reaction: An exactly solvable model}, Proc. Roy.
Soc. London A {\bf 442}, 641-658  and {\it Chaotic Classical and 
Half-classical Adiabatic Reactions: Geometric Magnetism and Deterministic 
Friction}, Proc. Roy. Soc. London A, {\bf 442}, 659-672.

\bibitem[Bersuker 1984]{bersuk} I.B. Bersuker, {\em The Jahn-Teller 
Effect and Vibronic Interactions in Modern Chemistry}, Plenum, New York.

\bibitem[Cohen-Tannoudji et al. 1969]{Cohen-T} C. Cohen-Tannoudji, J. 
DuPont-Roc, S. Haroche and F. Lalo\"e, {\em Detection of the static 
magnetic field produced by the oriented nuclei of optically pumped $^3$He 
gas}, Phys. Rev. Lett., {\bf 22}, 758-760.

\bibitem[Delacr\'etaz et al. 1986]{delcatraz} G. Delacr\'etaz, E.R. Grant, 
R.L. Whetten, L. W\"oste and J.W. Zwanziger, {\em Fractional quantization of
molecular pseudorotation in $Na_3$}, Phys.\ Rev.\ Lett.\ {\bf 56}, 2598-2601.

\bibitem[Dubrovin and Novikov 1980]{dn} B.A.~Dubrovin and S.P.~Novikov, {\it
Ground state of a two dimensional electron in a periodic magnetic field}, Sov.\
Phys.\ JETP  {\bf 52}, 511-516; {\it Ground State in a Periodic
Field, Magnetic Bloch Functions and Vector Bundles}, Sov.\ Math.\ Dokl.\ {\bf
22}, 240-244.

\bibitem[Englman 1972]{englman} R.~Englman, {\em The Jahn-Teller Effect in
Molecules and Crystals}, Wiley-Interscience, London.

\bibitem[Fr\"ohlich and Studer 1993]{fs} J.~Fr\"ohlich and U.~Studer, {\it Gauge
invariance and current algebras is non-relativistic many body theory}, Rev.\
Mod.\ Phys.\ {\bf 65}, 733-802.

\bibitem[Grossmann 1972]{grossmann} A.~Grossmann, {\it Momentum-like constants
of motion}, Statistical Mechanics and Field Theory, R.N.\ Sen and C.\ Weil Ed.\
Halsted Press.

\bibitem[Ham 1987]{Ham} F.S. Ham, {\it Berry's Geometrical Phase and the
sequence of states in the Jahn Teller effect}, \prl {\bf 58} 725-728.
 
\bibitem[Kato 1958]{kato58} T.~Kato, {\it On the adiabatic theorem of quantum
mechanics}, Phys.\ Soc.\ Jap.\ {\bf 5}, 435-439.
 
\bibitem[Kato 1966]{kato} T.~Kato, {\it Perturbation Theory of Linear 
Operators}, Springer.

\bibitem[King-Smith and Vanderbilt 1993]{vanderbilt} R.D.~King-Smith and
D.~Vanderbilt, {\em Theory of polarization of crystalline solids}
Phys.\ Rev.\ B {\bf 47} 1651-1654.

\bibitem[Klein and Seiler 1989]{ks} M.~Klein and R.~Seiler, {\it Vanishing
of Corrections to the Kubo Formula for Adiabatic  Charge Transport in Quantum
Hall Systems}, Comm.\ Math.\  Phys.\ {\bf 128}, 141-160.

\bibitem[Kohmoto 1985]{kohmoto}M.~Kohmoto, {\it Topological invariants and the
Quantum Hall effect},  Ann.\ Phys.\ {\bf 160} 343.

\bibitem[Kreft and Seiler 1996]{kreft-seiler} Ch. Kreft and R. Seiler, 
{\em Models of the Hofstadter-type}, J. Math. Phys. {\bf 37}, 5207-5243.

\bibitem[Krichever 1983]{krichever} I.M.~Krichever, {\it Nonlinear Equations and
elliptic curves}, Itogi Nauki i Tekhniki, {\bf 23} 79-136.

\bibitem[Kunz 1993]{kunz} H.~Kunz, {\it Adiabatic charge transport and topological
invariants for electrons in quasi-periodic potential and magnetic
field}, Helv.\ Phys.\ Act.\ {\bf 66}, 263-335.

\bibitem[Longuet-Higgins 1975]{longuet}H.C.~Longuet-Higgins, {\it The
intersection of potential energy surface in polyatomic molecule}, Proc.\ R.\
Soc.\ London A {\bf 344}, 147-156.

\bibitem[Mead 1992]{mead} C.A. Mead, {\em The geometric phase in molecular
systems}, Rev. Mod. Phys. {\bf 64}, 51-85.

\bibitem[Mead and Truhlar 1979]{mt}C.A.~Mead and D.G.~Truhlar, {\em On the
determination of Born-Oppenheimer nuclear motion wave functions including
complications due to conical intersections and identical nuclei}, 
J.\ Chem.\ Phys.{\bf 70}, 2284-2296.

\bibitem[Nenciu 1993]{nenciu} G.~Nenciu, {\it Linear Adiabatic
theory: Exponential estimates}, Comm.\ Math.\ Phys.\ {\bf 152}, 479-496.

\bibitem[Niu 1986]{niu86} Q.~Niu {\it Quantum Adiabatic particle transport},
Phys.\ Rev.\ B {\bf 34}, 5093-5100.

\bibitem[Niu 1990]{niu90} Q.~Niu, {\it Towards a quantum pump of electric
charges}, Phys.\ Rev.\ Lett.\ {\bf 64} 1812-1815.

\bibitem[Niu 1991]{niu91} Q.~Niu, {\it Theory of the quantized adiabatic particle
transport}, Modern Phys.\ Let.\ B  {\bf 5} 923-931.

\bibitem[Niu 1994]{niu94} Q.~Niu {\it Quantum density response in insulators},
Phys.\ Rev.\ B {\bf49}, 13~554-13~559.

\bibitem[Novikov 1981]{novikov} S.P.~Novikov, {\it Magnetic Bloch functions and
vector bundles,
Typical dispersion laws and their quantum numbers},  Sov.\ Math.\ Dokl.\ {\bf
23}, 298-303.

\bibitem[O'Brien 1989]{obrien} M.C.M. O'Brien, {\it The Berry phase in a
$T\otimes\tau_2$ Jahn-Teller system with a note on Tunnelling}, J.Phys. A 
Math. Gen. {\bf 22}, 1779-1797.

\bibitem[Shilton et al. 1996]{pepper} J.M.~Shilton, V.I.~Talyanskii, M.~Pepper,
D.A.~Ritchie, J.E.F.~Frost, C.J.B.~Ford, C.G.~Smith and G.A.C.~Jones, {\it High
frequency single   electron transport in a quasi one dimensional GaAs channel
induced by surface acoustic waves}, J.\ Phys.: Cond.\ Matt.\ {\bf 8}
L531-L539.

\bibitem[Poilblanc et al. 1991]{montambaux} D.~Poilblanc, G.~Montambaux,
M.~Heritier and P.~Lederer, Phys.\ Rev.\ Lett. {\bf 58}, 270.

\bibitem[Peierls 1991]{peierls} R.~Peierls, {\em More Surprises in Theoretical
Physics}, Sec. 2.3, P.~U.~P., Princeton.

\bibitem[Purcell 1977]{purcel} E. Purcell, {\it Life at low Reynolds numbers},
Am.\ J.\ Phys.\  {\bf 45} 3.

\bibitem[Resta 1994]{resta} R.~Resta, Rev.\ Mod.\ Phys.\ {\bf 66}, 899.

\bibitem[Resta and Sorella 1995]{ressor} R. Resta  and S. Sorella, 
Phys.\ Rev.\ Lett. {\bf 74}, 4738.

\bibitem[Shapere and Wilczek 1989a]{sw} A.~Shapere and F.~Wilczek, {\it
Geometric Phases in Physics}, World Scientific, Singapore.

\bibitem[Shapere and Wilczek 1989b]{sw89} A.~Shapere and F.~Wilczek, {\it
Geometry of self propulsion at low
Reynolds number}, J.\ Fluid Mech.\ {\bf 198} 557-585.

\bibitem[Seiler 1991]{seiler91} R.~Seiler, {\it On the quantum Hall
effect}, in {\it Recent developments in Quantum Mechanics}, A.\ Boutet
de Monvel et al.\, Eds., Kluwer, Netherland.

\bibitem[Simon 1983]{simon83} B.~Simon, {\it Holonomy, the quantum
adiabatic theorem and Berry's phase},  Phys.\ Rev.\ Lett. {\bf 51}
2167-2170.

\bibitem[Stone 1992]{Stone} M.~Stone, {\it The Quantum Hall effect},
World Scientific, Singapore.

\bibitem[Tan 1994]{tan} Y. Tan {\it Localization and quantum Hall effect in two
dimensional periodic potential}, J.\ Phys.\ C {\bf 6}, 7941-7954.

\bibitem[Teller 1937]{jahnteller} E.~Teller, J.\ Phys.\ Chem.\ {\bf 41} 109. 

\bibitem[Thouless 1983]{thouless83} D.J.~Thouless, {\it Quantization of particle
transport}, Phys.\ Rev.\ B {\bf 27}, 6083.

\bibitem[Thouless 1994]{thouless94} D.J.~Thouless,{\it Topological interpretation
of quantum Hall conductance}, J.\ Math.\ Phys.\ {\bf 35}, 1-11.

\bibitem[Thouless et al. 1982]{tknn} D.~J.~Thouless, M.~Kohmoto,~P.~Nightingale and
M. den Nijs, {\it Quantum Hall conductance in a two dimensional periodic
potential}, Phys.\ Rev.\ Lett. {\bf 49}, 40.

\bibitem[Whetten et al. 1986]{st} R.L. Whetten K.S. Haber and E.R. Grant, 
J. Chem. Phys. {\bf 84}, 1270.

\bibitem[Wigner 1959]{wigner} E.P.~Wigner, {\it Group Theory}, Academic, N.Y.

\bibitem[Wilkinson 1984]{wilkinson} M.~Wilkinson, {\it An example of phase holonomy
in WKB theory}, J.\  Phys.\  A {\bf 17}, 3459-3476.

\bibitem[Yakovenko 1991]{yano}
V.M.Yakovenko, Phys.\ Rev.\ B {\bf 43}, 11~353-11~366.

\end{thebibliography}
\end{document}